\documentclass{article}
\usepackage{iclr2026_conference,times}

\usepackage{amsmath,amsfonts,bm}

\def\eqref#1{equation~\ref{#1}}

\def\1{\bm{1}}

\DeclareMathAlphabet{\mathsfit}{\encodingdefault}{\sfdefault}{m}{sl}
\SetMathAlphabet{\mathsfit}{bold}{\encodingdefault}{\sfdefault}{bx}{n}

\usepackage{hyperref}
\usepackage{url}
\usepackage{booktabs}
\usepackage{graphicx}
\usepackage{makecell}
\usepackage[scaled=0.85]{FiraMono}

\title{Does Code Cleanliness Affect Coding Agents? \\ A Controlled Minimal-Pair Study}

\author{Priyansh Trivedi\\
SonarSource \\
\texttt{priyansh.trivedi@sonarsource.com}
\AND
Olivier Schmitt\\
SonarSource \\
\texttt{olivier.schmitt@sonarsource.com}}

\iclrfinalcopy

\begin{document}
\maketitle

\lhead{}

\begin{abstract}
As autonomous coding agents see rapid adoption, their evaluation has primarily focused on task completion rates holding the target codebase fixed.
This leaves a critical question unanswered: does the structural and stylistic quality, or ``\textit{cleanliness}'' of the underlying code affect an agent's ability to navigate and modify it?
To isolate the effect of code cleanliness from agent capability, we introduce an evaluation protocol built around \textit{minimal pairs}: repositories that match on architecture, dependencies, and external behaviour, but differ on static-analysis rule violations and cognitive complexity.
The pairs are constructed in both directions, by agent pipelines that either degrade a clean repository or clean a messy one. We author 33 tasks across six such pairs, evaluated through hidden tests at the application's public surface. 
Across 660 trials with Claude Code, code cleanliness does not change the agent's pass rate.
However, it substantially alters the agent's operational footprint: agents working on cleaner code use 7 to 8\% fewer tokens and reduce file revisitations by 34\%. 
Our findings suggest that traditional maintainability principles remain highly relevant in the era of AI-driven development, shaping the computational cost and navigational efficiency of coding agents.
Code cleanliness joins model choice, harness, and prompting as a factor that materially affects agent behaviours. %

\end{abstract}

\section{Introduction}
\label{sec:intro}

Autonomous \textit{coding agents} are seeing rapid adoption and are reshaping how software gets written. A 2026 survey of 128{,}018 GitHub projects finds traces of agent activity in 22--29\% of them, in codebases of all sizes and ages, less than a year after the first practical agents shipped \citep{robbes2026agenticmuch}. Running these agents is not cheap. They accumulate a large \textit{token footprint} per task: on SWE-bench Verified, a single task averages around 4 million tokens across frontier LLMs, with input tokens accounting for most of that~\citep{bai2026, wang2025agenttaxo, salim2026}.

Evaluation of coding agents has focused primarily on whether they solve the task: pass-rate on benchmarks such as SWE-bench~\citep{jimenez2024swebench, chowdhury2024swebenchverified}. A more recent thread reports the resources an agent consumes alongside its pass rate~\citep{fan2025sweeffi, bai2026, wang2025agenttaxo, salim2026}, finding among other things that token usage is highly variable across models and runs, and that input tokens dominate the total. Common to this thread is that the codebase is held fixed; what varies is the agent or its scaffolding. To our knowledge, no work has held the agent and the task fixed, and varied the codebase.

What we mean by ``the codebase'' here is a particular property of it: \textit{cleanliness}. The bundle of structural and stylistic characteristics associated with maintainable code: readability, low cognitive complexity, factored helpers, clear naming. A 2,000-line method with deep nesting is plausibly more expensive for an agent to navigate than the same behavior factored across small named helpers. A predictably named method like \texttt{normalize\_query} is plausibly easier to find than \texttt{\_xfm\_q2}.

We organize the study around the following questions:

\begin{itemize}
    \item \textbf{RQ1.} Does code cleanliness impact coding agents' ability to successfully complete tasks?
    \item \textbf{RQ2.} Does it change the operational footprint the agent leaves doing that work --- input and output tokens, reasoning, conversation length, files read, file revisitations, lines edited?
    \item \textbf{RQ3.} Does the effect vary with task topology, particularly between work concentrated in a single dense code region and work spanning multiple module boundaries?
\end{itemize}

To answer these, we need a comparison the open-source ecosystem does not provide: two repositories identical on architecture, dependencies, tests, and external behavior, differing on cleanliness alone. Such pairs do not arise naturally, and they are not present in any benchmark we are aware of. The comparison apparatus has to be built before the question can be asked.

We therefore construct \textit{minimal pairs}: pairs of repositories that share architecture, dependencies, tests, and external behavior, but differ on cleanliness. 
We use SonarQube rule violations as a proxy for cleanliness (detailed in Section~\ref{sec:benchmark}).
A pair of repositories on its own is inert: it gives us two states to compare, but no occasion on which to compare them. 
We therefore also author 33 coding tasks across the pairs, shaped like instances in existing agentic benchmarks. %
Section~\ref{sec:benchmark} details the construction.

Our contributions can be summarized as:

\begin{itemize}
    \item A protocol for controlled measurement of code cleanliness effects on agent's token footprint, built around bidirectional minimal-pair construction and hidden tests at the public surface.
    \item An actionable benchmark of six minimal-pair repositories, plus 33 tasks with reference implementations, and hidden tests. %
    \item On Claude Code (using Claude Sonnet 4.6), we find that the agents' ability to solve a task does not change based on code cleanliness. However, agents working on cleaner code have a smaller footprint: by 7 to 8\% on token-equivalent metrics and around 34\% on file revisitation. %
\end{itemize}

\section{Benchmark Construction}
\label{sec:benchmark}

Concretely, this benchmark consists of (i) a set of \textit{minimal pairs} of code repositories, where the two repositories are behaviourally equivalent, but differ only on cleanliness, and (ii) a task suite that exercises that difference directly.

The benchmark is built on the Harbor framework\footnote{v0.4.0}~\citep{Harbor_Framework}, and both halves are designed for reuse so that other models, other harnesses, and downstream questions about how the codebase shapes the agent can run against the same apparatus.
\textit{Clean code} is not a precisely-defined property, and we do not advance a formal definition.
We use the term for a family of characteristics widely associated with maintainable code: small, single-purpose functions; names that describe what those functions do; control flow a reader can follow without effort; little dead code, duplicated logic, or accidental coupling between unrelated parts of the system.
In our work, we loosely quantify cleanliness by using a static code analyzer -- SonarQube\footnote{\href{https://www.sonarsource.com/products/sonarqube/cloud/}{SonarQube Cloud}, Enterprise Edition. Rule set: ``default quality gate''}: code is \textit{cleaner} to the extent that it carries fewer rule violations (interchangeably referred to as issues).

We say two repositories are \emph{behaviourally equivalent} when they produce the same externally observable outputs (input-output mappings and state transitions) under the same inputs, regardless of how their code is internally organized. Operationally, we treat the two sides of a pair as behaviourally equivalent when they pass the same test suite at equal coverage, or a pair of test suites slightly modified to compensate for non-observable code changes such as refactoring.

\subsection{Constructing minimal pairs}
\label{sec:minimal_pairs}

In our experiments, we measure how an agent behaves on a task across two versions of an application, and we want to attribute any observed difference in its behaviour to the state of the code itself.
That reading is only valid if the two versions match on everything else that plausibly shapes the agent: same language, same framework, same dependencies, same external behaviour.
This section, therefore, outlines how such pairs are built, for our benchmark. 
Each pair is constructed in one of two directions by an agent-driven pipeline that degrades or cleans a repository without altering the code's behaviour.
Each pipeline aims to produce an alternate-history version of the codebase it starts from.
The degradation pipeline takes a codebase that has historically utilized static analysis to filter out rule violations, and generates a version that likely reflects what the codebase might have looked like without it.
The cleanup pipeline approaches from the opposite direction, taking a codebase with a naturally high volume of rule violations, and producing what it would resemble had it been actively gated by an automated analyzer. 
The two directions produce pairs of slightly different shapes, which we return to below.

\textbf{Degrading a clean repository (Slopify).}
The pipeline takes a clean codebase and produces a degraded version of it: code that plausibly grew on a team without code review or linting, not code that was deliberately sabotaged.
The pipeline runs in three phases, each dispatched to a fresh agent.
The three-step process was arrived at to minimize the time and derailments coding agents undergo when asked to tackle the problem from scratch on a given directory of a code repository:
\begin{itemize}
  \item \textit{Build.} Get the repository compiling and its test suite passing, and freeze the working commands into a \texttt{build\_instructions.md} at the repository root.
  \item \textit{Explore.} Walk the repository and write a \texttt{summary.md} in each directory worth degrading. %
  \item \textit{Transform.} Iterate over the flagged directories, introducing issues from a curated set of SonarQube rules until the module is sufficiently degraded. Tests are rerun after each pass, and any pass that breaks them is rejected.
\end{itemize}

\textbf{Cleaning an unclean repository (Vibeclean).}
The pipeline takes an organically-grown codebase and resolves the issues present in it, with the application's external behaviour intact.
The agent's target list is the list of issues itself: each issue is anchored to a span of code, and the agent works through the list, editing that code until the issue is resolved.
The working scope is bounded by what the analyser can flag, and not an open mandate to redesign.

The pipeline runs in two phases, each dispatched to a fresh agent:
\begin{itemize}
  \item \textit{Build.} As in the degradation pipeline above: get the repository compiling and its test suite passing, and freeze the working commands into a \texttt{build\_instructions.md}.
  \item \textit{Clean.} The agent works module by module, and mechanically cleans up relevant analyzer rule violations. Once done with a module, tests are run to ensure behavioural parity. %
\end{itemize}

Across both pipelines we assembled six minimal pairs (Table~\ref{tab:pairs}) of code repositories.
They split into three primarily Java and three primarily Python codebases, with some carrying a small amount of code in the other language.
Three are drawn from public, open-sourced repositories; three from private SonarSource codebases.
The private pairs guard against memorisation: the LLM under evaluation has plausibly trained on the public repositories, but it cannot have trained on the private ones.

The transformations made by the slopify pipeline, like inlining helpers into their callers, duplicating logic across paths, padding files with dead code, occasionally merging modules into single files, all add cognitive complexity that wasn't there before.
The edits made by the vibeclean pipeline are of a different shape.  
It deduplicates string literals, deletes commented-out code, replaces legacy collection idioms, and removes dead branches. 
The cleanup agent also performs structural rewrites where the analyser flags genuine god structures, such as replacing 200-plus-line dispatch switches with named helpers, or extracting persistence helpers out of a 2,800-line class. 
But extraction redistributes complexity across more methods rather than eliminating it, and a handful of the largest god structures are declined as wontfix. 

\begin{table}[h]
\centering
\caption{The six minimal pairs used in our experiments. Repositories marked with ${}^{*}$ are private codebases; the rest are public. Each metric column reports the value on the cleaner side / messier side of the pair. \textit{NCLOC} is the non-comment line count, in thousands. \textit{Issues} is the SonarQube issue count. \textit{Iss/kLOC} and \textit{CogComp/kLOC} are issue density and cognitive-complexity density, both per thousand NCLOC. NCLOC is computed as issue count divided by issue density, rounded to one decimal.}
\label{tab:pairs}
\resizebox{\textwidth}{!}{%
\begin{tabular}{lccccc}
\toprule
Repository & Pipeline & NCLOC (k) & Issues & Iss/kLOC & CogComp/kLOC \\
\midrule
\texttt{sonar-sca}${}^{*}$           & Slopify   & 128.8 / 136.7 & 94 / 2{,}825    & 0.73 / 20.66 & 30.6 / 56.5   \\
\texttt{sonar-caas-poc}${}^{*}$      & Slopify   & 26.2 / 31.5   & 16 / 855        & 0.61 / 27.16 & 179.8 / 218.9 \\
\texttt{sonarcloud-codedatalake}${}^{*}$ & Slopify & 45.6 / 38.4   & 199 / 1{,}319   & 4.36 / 34.39 & 34.0 / 216.5  \\
\href{https://github.com/apache/commons-bcel}{\texttt{commons-bcel}}                & Vibeclean & 55.1 / 54.8   & 694 / 2{,}711   & 12.60 / 49.46 & 102.8 / 108.3 \\
\href{https://github.com/Netflix/genie}{\texttt{genie}}                       & Vibeclean & 118.8 / 116.7 & 152 / 1{,}262   & 1.28 / 10.81 & 22.2 / 23.5   \\
\href{https://github.com/ckan/ckan}{\texttt{ckan}}                        & Vibeclean & 133.4 / 132.1 & 1{,}006 / 3{,}632 & 7.54 / 27.50 & 69.3 / 76.5   \\
\bottomrule
\end{tabular}%
}
\end{table}

\subsection{Designing tasks}
\label{sec:tasks}

Whether cleanliness shows up in an agent's behaviour (correctness, token footprint) depends on the parts of the codebase the task requires reading/editing.
If a task lives in a part of the codebase where the two sides of a pair look the same, the agent never encounters the difference we care about.
If a task description names the files or functions to edit, the agent skips the exploration phase, plausibly the part where cleanliness costs the most.
If tests check internal structure rather than external behaviour, the cleaner and messier versions of the pair will pass them differently for reasons unrelated to cleanliness.
The way each task is designed therefore decides what the experiment can measure.

Three rules follow from this, and every task in the benchmark obeys them:
\begin{itemize}
  \item \textit{Route through hotspots.} Each task is anchored to code regions where the two sides of the pair differ most in issue density and cognitive complexity.
  \item \textit{Describe in externally observable terms.} A task's description gives the inputs and outputs the implementation must support, along with example scenarios. It names no files, functions, or other internal structure; the agent does the exploration itself.
  \item \textit{Test at the public surface.} Each task's hidden tests interact with the application through whatever interface it presents to its callers (a CLI, HTTP routes, a library or framework API, or the repository's equivalent).
\end{itemize}

\begin{figure}[t]
\centering
\includegraphics[width=\linewidth]{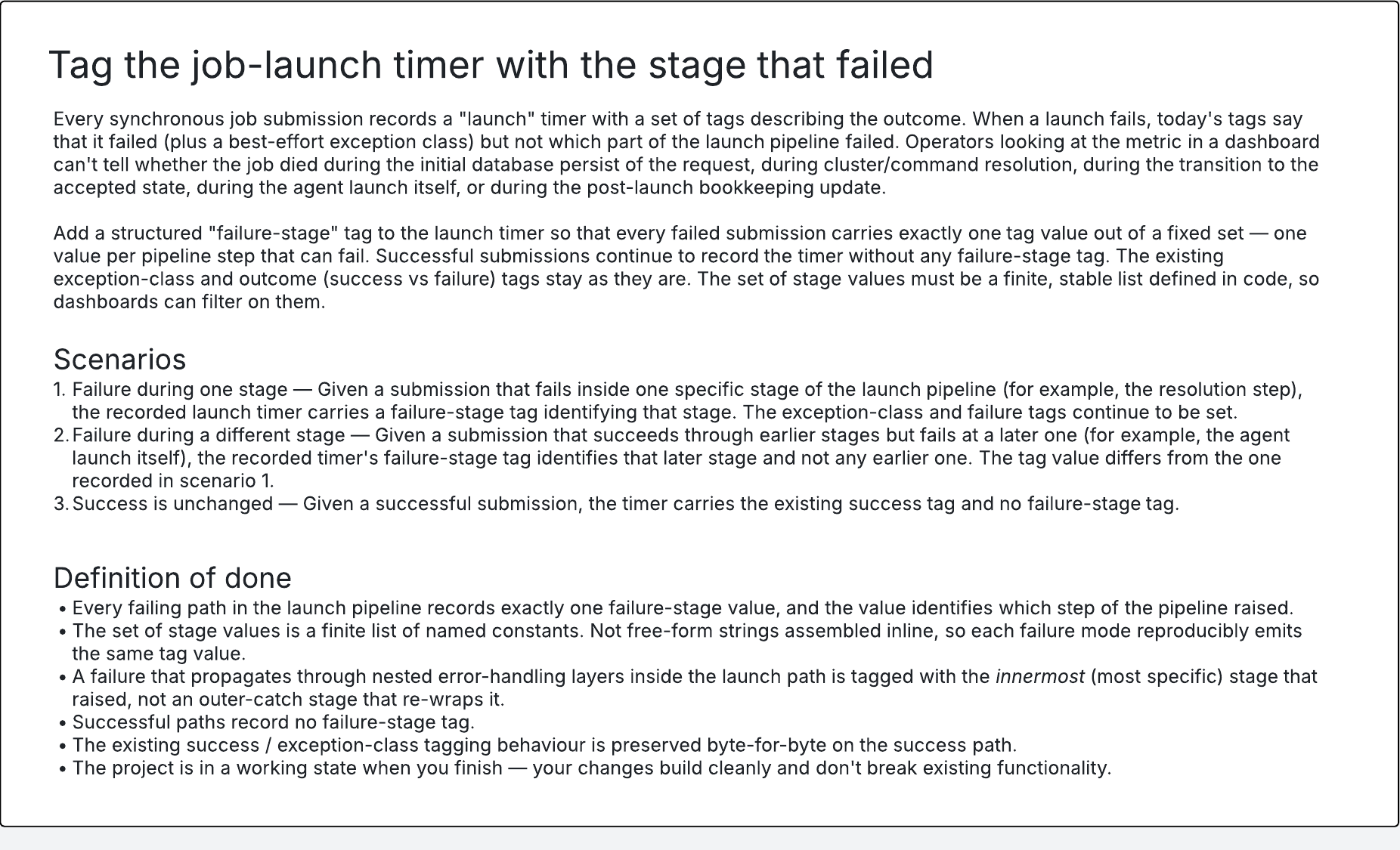}
\caption{An example task in the benchmark, drawn from the \texttt{genie} pair. The agent reads an externally observable description (shown) and produces a code change that a hidden test suite, kept internal, exercises against the application's public surface. This task asks the agent to add a structured failure-stage tag to Genie's synchronous job-launch timer so that dashboards can attribute job-launch failures to a specific pipeline stage.}
\label{fig:task-example}
\end{figure}

The actual task generation process has been split into five phases to compensate for the aforementioned challenges. In the first phase, agents are tasked with comparing the cleaner and messier variants of a repository and creating a map of differences. Armed with this, a second agent creates task-outlines including what parts of the repositories may be upstream, or downstream of expected changes, and how we may test for them. We then manually select, edit and curate the most plausible and interesting outlines. A third agent then converts them into actionable tasks: the actual instruction the experimental agent will read, a hidden test suite at the application's public surface, and a reference implementation that we keep internal.
Finally, two automated checks are run on every authored task before it enters the dataset.
The reference implementation must pass the aforementioned hidden tests on both sides; if it does not, the tests are not describing the behaviour we asked for.
The untouched repository must not pass the hidden tests; if it does, the tests are not measuring the change.
Tasks that fail either check after two iterations are rewritten by hand, or dropped.

Writing reference implementations during human review helped eliminate tasks that read well as outlines but were practically too simple or too complex.
The reference implementation itself is not required for running the benchmark; the value of writing it sat largely in this sanity check.
An example of one such task, authored on \texttt{genie}, is shown in Figure~\ref{fig:task-example}.

The final benchmark contains 33 tasks across the six minimal pairs of repositories. %
They're divided in three \textit{tracks}, each probing a different mechanism by which cleanliness might matter:
\begin{itemize}
  \item 13 \textbf{Cognitive-hotspot tasks.} Each task is routed through a region of dense single-method or single-class complexity. These probe whether navigating a god-method or a deeply-nested control flow costs the agent more.
  \item 14 \textbf{Multi-module tasks.} Each task requires changes spanning two or more modules. These probe whether crossing seams is more expensive when the seams are messy.
  \item 6 \textbf{Calibration tasks.} Each task is built to be cleanliness-insensitive: simple work in regions where the two sides of a pair are identical (for example, adding a log line at CLI entry). They tell us whether anything is moving between sides when we would expect nothing to.
\end{itemize}

\section{Experimental Setup}
\label{sec:setup}

All experiments use Claude Code as the agent, with its default tool set.
The agent reads only the task description; it receives no additional priming about cleanliness, and is blind to which side of a pair it is working on.
All numbers reported in this paper come from Claude Sonnet 4.6 trials.
We discuss the limits of measuring under a single configuration in Sec.~\ref{sec:limitations}.
We also swept Claude Haiku 4.5 over the same task set; its pass rate was too low to read footprint differences cleanly, so we exclude it from Sec.~\ref{sec:results}.
Each task is run ten times per side of its pair, producing $33 \times 2 \times 10 = 660$ trials in total.

Each trial runs in a containerised sandbox with bounded CPU, memory, storage, and wall-clock budgets, plus access to public package registries.
The base image is repo-specific, carrying the toolchain, build caches, and services each codebase needs to compile and run its tests.
Within each pair, the two sandboxes are identical except for the source tree mounted at \texttt{/app}.

For each trial we record ten metrics, grouped by what they say about the agent's behaviour on the task.

\textit{Pass rate} is the fraction of the task's hidden tests that pass on the agent's final state.
This is the metric we would have if we cared only whether the agent got the work done.

The three \textbf{footprint} metrics measure the size of the conversation the agent produces.
\textit{Input tokens} count every token the model reads across all turns of a trial, dominated by file contents pulled into context and earlier conversation re-sent on each turn. This also includes any tokens consumed by a subagent that may be launched in the process of executing the task. We do not distinguish between cached and uncached tokens since aspects beyond the code,  such as wall-time,
 might influence them.
\textit{Output tokens} measure everything emitted by the model and any sub-agents: prose, code, reasoning traces, and tool invocations. While it might be tempting to consider output tokens as a proxy for "lines written" by the agent, note that each tool call that the agent invokes also involves the LLM \textit{outputting} the tool name and arguments. 
\textit{Reasoning characters.} The Anthropic API folds reasoning tokens into the output-token count and does not surface them separately. In its absence we count the plaintext characters in the reasoning-content blocks of each trajectory. Reasoning characters overlap with output tokens by construction (reasoning is already counted in there). We read them as an additional signal alongside the token counts, not as an independent footprint dimension.

The three \textbf{trajectory} metrics describe how the conversation unfolds over time.
\textit{Conversation turns} are the total number of agent-tool exchanges in a trial.
\textit{Turns before first edit} are the turns elapsed before the agent issues its first file modification.
\textit{Characters before first edit} are the total conversation characters elapsed in that same window.
The last two capture how much the agent explores before committing to a change.

The three \textbf{navigation} metrics capture how the agent moves through and modifies the codebase.
\textit{Files read} is the number of distinct files the agent opens during a trial.
\textit{File revisitation} counts how often the agent re-reads a file it has already read and edited; a typical revisitation in a trajectory looks like \textit{read}~$\to$~\textit{edit}~$\to$~(possibly other work)~$\to$~\textit{read again}.
We interpret revisitation as uncertainty about a previous edit rather than as breadth of exploration.
\textit{Lines edited} is the number of source lines the agent modifies during the trial, the size of the final patch.

Agent footprint metrics are known to vary substantially across repeated runs of the same task, even at fixed temperature \citep{bai2026}.
We therefore apply a median-threshold \textit{outlier filter} within each (task, side) combination: trials more than $50\%$ off the median of its ten repetitions are dropped before averaging. In practice this drops $9.7\%$ of trials.
Dataset-level numbers are micro-averaged across the 33 tasks: for each metric $m$, per-task means on each side are summed, and the percent difference is computed once over the pooled totals,
\begin{equation}
\label{eq:delta}
\Delta m \;=\; \frac{\sum_{\text{task}} \bar{m}^{\text{cleaner}}_{\text{task}} \;-\; \sum_{\text{task}} \bar{m}^{\text{messier}}_{\text{task}}}{\sum_{\text{task}} \bar{m}^{\text{messier}}_{\text{task}}},
\end{equation}
where $\bar{m}^{\text{cleaner}}_{\text{task}}$ is the mean of metric $m$ over the \texttt{cleaner}-side trials of a task. %
Every percentage difference reported in the paper follows this convention, except for pass rate, for which we report the absolute difference in percentage point ($\bar p^{\text{cleaner}}-\bar p^{\text{messier}}$).

\section{Results}
\label{sec:results}

\begin{table}[h]
\centering
\caption{Per-repo aggregate effects, micro-averaged within each repo, and across the 33 tasks of the benchmark for the bottom row. 
All cells use the outlier filter from Sec.~\ref{sec:setup}.
\textit{Msgs $\to$ 1st} and \textit{Chars $\to$ 1st} are the ``Turns before first edit'', and ``Characters before first edit'' metric, outlined amongst others in Sec.~\ref{sec:setup}.
}
\label{tab:results}
\scriptsize
\setlength{\tabcolsep}{3pt}
\resizebox{\textwidth}{!}{%
\begin{tabular}{lrrrrrrrrrr}
\toprule
Repo & \shortstack{Pass\\rate} & \shortstack{Input\\tok} & \shortstack{Output\\tok} & \shortstack{Rsng\\chars} & \shortstack{Conv.\\msgs} & \shortstack{Msgs\\$\to$ 1st} & \shortstack{Chars\\$\to$ 1st} & \shortstack{Files\\read} & Revisits & \shortstack{Lines\\edited} \\
\midrule
ckan                            & $+5.0$\,pp & $+0.8\%$   & $-11.4\%$ & $-17.8\%$ & $-5.3\%$  & $+17.5\%$ & $+25.2\%$ & $-3.0\%$  & $-7.4\%$  & $-13.9\%$ \\
commons-bcel                    & $-6.5$\,pp & $-17.4\%$  & $-16.9\%$ & $-13.2\%$ & $-16.1\%$ & $-10.0\%$ & $-5.6\%$  & $+7.1\%$  & $-68.5\%$ & $-22.4\%$ \\
genie                           & $-3.7$\,pp & $+5.8\%$   & $-9.1\%$  & $-4.5\%$  & $+0.7\%$  & $+1.2\%$  & $+8.9\%$  & $+5.0\%$  & $-25.5\%$ & $-12.4\%$ \\
sonar-caas-poc                  & $-3.3$\,pp & $-0.5\%$   & $+1.7\%$  & $-3.2\%$  & $+1.1\%$  & $-1.1\%$  & $-3.5\%$  & $+10.9\%$ & $-34.2\%$ & $+16.3\%$ \\
sonar-sca                       & $+1.5$\,pp & $-4.7\%$   & $+1.0\%$  & $+0.1\%$  & $-0.4\%$  & $-3.1\%$  & $-3.0\%$  & $+4.3\%$  & $-28.6\%$ & $-3.5\%$  \\
\makecell[tl]{sonarcloud-\\codedatalake} & $+1.3$\,pp & $-29.0\%$  & $-13.5\%$ & $-16.6\%$ & $-23.1\%$ & $-19.2\%$ & $-20.6\%$ & $-7.6\%$  & $-48.6\%$ & $+0.0\%$  \\
\midrule
\textbf{Dataset} & $\mathbf{-0.9}$\,\textbf{pp} & $\mathbf{-7.1\%}$ & $\mathbf{-8.5\%}$ & $\mathbf{-11.1\%}$ & $\mathbf{-7.0\%}$ & $\mathbf{-3.6\%}$ & $\mathbf{-4.6\%}$ & $\mathbf{+3.2\%}$ & $\mathbf{-33.8\%}$ & $\mathbf{-3.2\%}$ \\
\bottomrule
\end{tabular}%
}
\end{table}

\subsection{Aggregate findings}
\label{sec:aggregate}

Cleanliness changes the agent's footprint and behaviour without changing whether the task gets done.
Table~\ref{tab:results} reports the per-repo effect of cleanliness on each of the ten metrics defined in Sec.~\ref{sec:setup}, with the dataset-level aggregate in the last row.
To begin with, the agents seem to accomplish the given tasks consistently, regardless of whether working on cleaner or messier code, as the pass rate changes by less than a percentage point between sides ($0.913$ on cleaner code, $0.921$ on messier code $=-0.9$~pp).

Footprint metrics shrink modestly but consistently on the cleaner side.
Input tokens fall $7.1\%$: the agent pulls less code into context.
Output tokens fall $8.5\%$, capturing the size of everything the model produces (tool calls, code, prose, and reasoning).
Reasoning characters, a subset of output tokens, drop $11.1\%$. 
Taken together with unchanged pass rate, this suggests the agent is doing the same work with fewer tokens. 

Trajectory metrics shift in the same direction, although less strongly.
Conversation messages drop $7.0\%$, in step with the footprint shrinkage.
The two metrics that measure how soon the agent commits to a first change (messages before first edit at $-3.6\%$, characters before first edit at $-4.6\%$) tend to favor cleaner code as well. We do not draw any inference from them alone, given their small magnitude. %

The largest single effect is on file revisitation: on the cleaner side, the agent returns to files it has already edited $34\%$ less often.
The direction holds on every repo, ranging from $-7\%$ on \texttt{ckan} to $-69\%$ on \texttt{commons-bcel}.
Files read goes up slightly at the dataset level ($+3.2\%$), though per-repo the direction is mixed and the magnitude small.
Taking the files-read change tentatively, a plausible reading is that on cleaner code the agent reads wider on its first pass and then commits to its changes, while on messier code it works through fewer distinct files but keeps returning to them to verify previous edits.
Lines edited shifts $-3.2\%$ at the dataset level with wide per-repo spread; we read this, like the small-magnitude trajectory metrics, as suggestive rather than independently informative.

\subsection{Per-track token footprint}
\label{sec:tracks}

The aggregate effects in Sec.~\ref{sec:aggregate} hide the differences in behaviour of agents when solving tasks of a specific track (as outlined in Sec.~\ref{sec:tasks}). Table~\ref{tab:tracks} reports the dataset-level deltas restricted to each track for the metrics driving the aggregate results, with the calibration track included as a neutral control.

\begin{table}[h]
\centering
\caption{Per-track aggregate effects, micro-averaged across the tasks in each track. Each cell is $(\text{cleaner}-\text{messier})/\text{messier}$ except pass rate, which is in percentage points. Filter stack as in Sec.~\ref{sec:setup} (outlier filter, subagent-inclusive metrics).}
\label{tab:tracks}
\begin{tabular}{lrrrrrr}
\toprule
Track & $n$ & \shortstack{Pass\\rate} & \shortstack{Input\\tok} & \shortstack{Files\\read} & Revisits & \shortstack{Lines\\edited} \\
\midrule
Cognitive-hotspot   & 13 & $+0.1$\,pp & $+1.8\%$  & $\mathbf{+11.2\%}$ & $-20.2\%$ & $-9.3\%$  \\
Multi-module        & 14 & $-2.6$\,pp & $\mathbf{-10.7\%}$ & $-0.6\%$  & $\mathbf{-50.8\%}$ & $+2.2\%$  \\
Calibration         & 6  & $\phantom{+}0.0$\,pp & $+3.8\%$  & $+2.0\%$  & $-75.0\%$ & $+12.0\%$ \\
\bottomrule
\end{tabular}
\end{table}

\textbf{Calibration track:}
The calibration row offers a sanity check on the procedure: pass rate is flat, and the token and files-read deltas sit within a few percent of zero. The outlying $-75\%$ revisitation delta comes down to three re-reads on a single task out of six, looking large only because we micro-average the numbers.%

\textbf{Multi-module track:}
On the fourteen multi-module tasks, the cleaner variant carries the experiment's strongest footprint contrast. The agent uses noticeably fewer input tokens, and file revisitations come down to roughly half their messier-variant level. The count of distinct files read is essentially unchanged across the two variants: when solving these tasks, the agent opens roughly the same number of files on both, but on the cleaner variant it commits to an edit and moves on, while on the messier variant it loops back to files it has already touched. Most of the dataset-level footprint reduction reported in Sec.~\ref{sec:aggregate} comes from this track.

\textbf{Cognitive-hotspot track:}
The thirteen hotspot tasks behave differently. Token footprint at this track level is essentially unchanged between the two variants. What changes is the shape of the work: on the cleaner variant the agent opens more files and edits fewer lines per file, while still reaching the same pass rate. This is consistent with how the cleanup pipelines transform hotspots in Sec.~\ref{sec:minimal_pairs}: extraction redistributes complexity across more methods rather than eliminating it, so the cleaner variant of a hotspot has its logic spread across more, smaller files for the agent to navigate through.

\textbf{A tension underneath the aggregate.}
The two non-calibration tracks pull in opposite directions on the per-task token footprint. 
A task that spans module boundaries shrinks the agent's footprint substantially on the cleaner side.
In contrast, a task whose hotspot has been refactored into helpers across files nets out at roughly neutral. Agents on the cleaner variant may have to reason less, but end up reading and patching more locations.
The $-7.1\%$ aggregate on input tokens is what remains once these two behaviours are averaged together. The per-task spread we examine next (Sec.~\ref{sec:variance}) is what the tension looks like one task at a time.

\subsection{Variance and per-task spread}
\label{sec:variance}

\begin{figure}[t]
\centering
\includegraphics[width=0.5\linewidth]{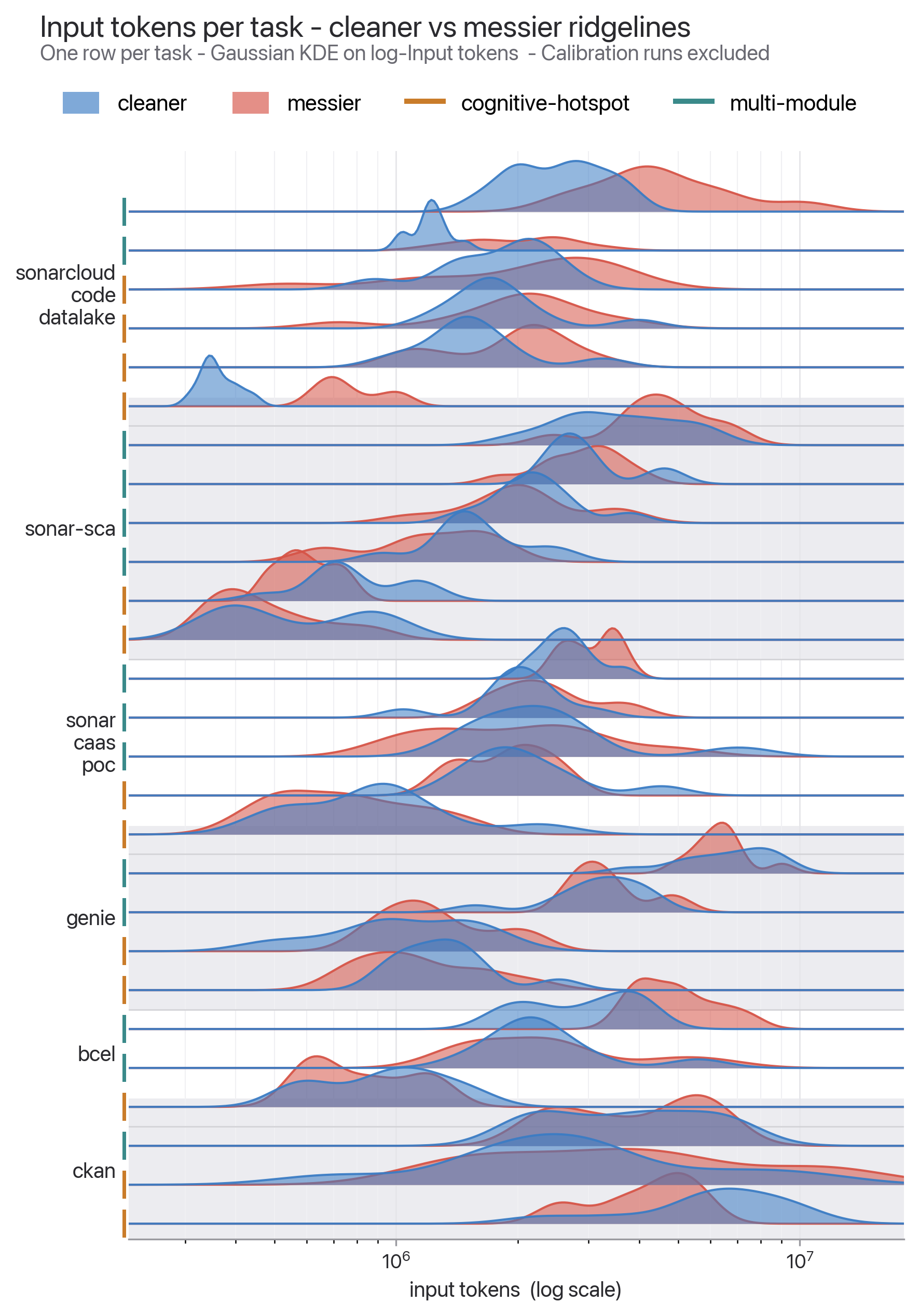}%
\includegraphics[width=0.5\linewidth]{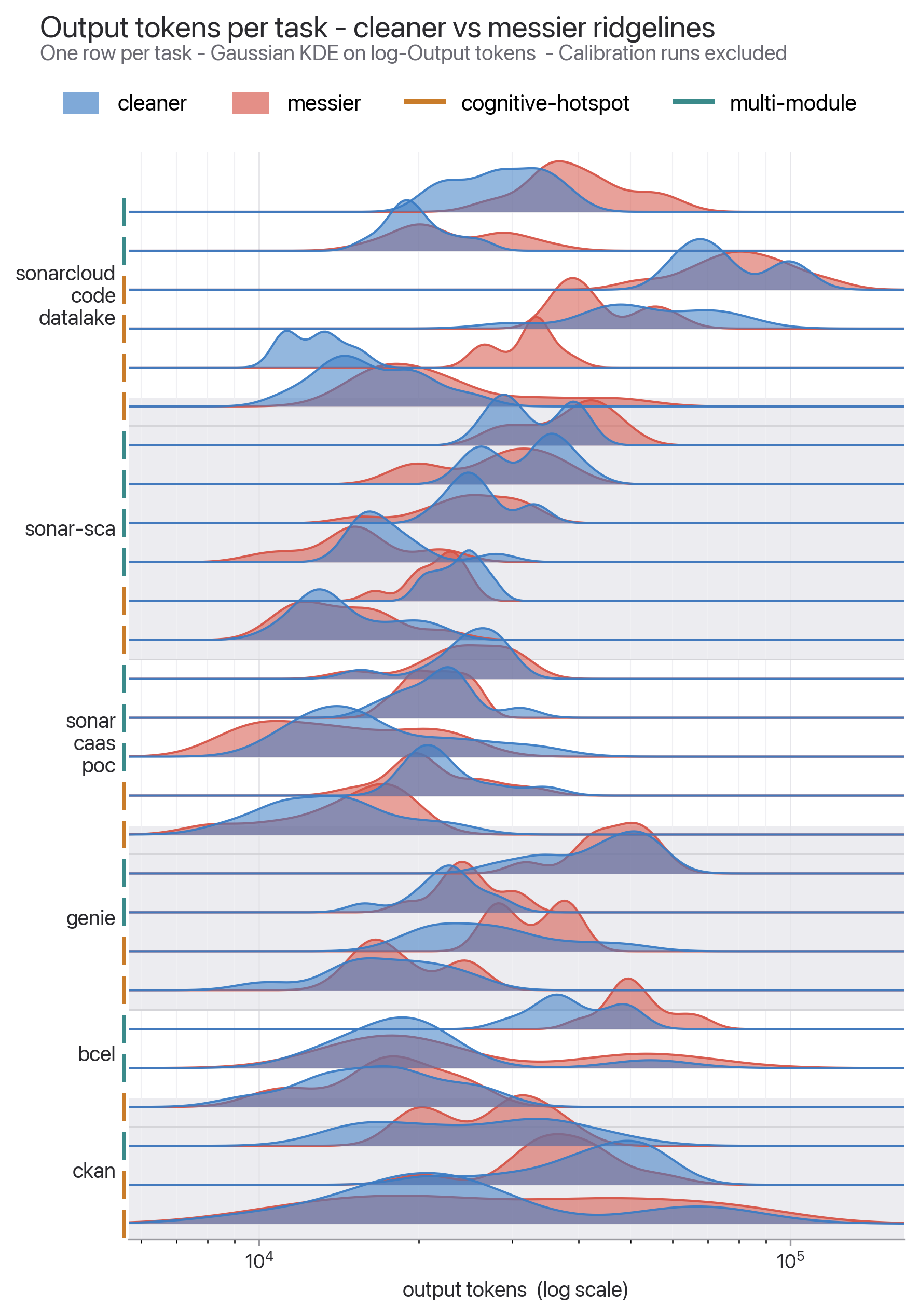}%
\caption{Per-trial distributions of input tokens (left) and output tokens (right) across the 27 tasks. Calibration tasks are excluded for clarity. Each row is a task, plotted as a Gaussian KDE on the log of the metric, with cleaner-side and messier-side trials overlaid. Calibration tasks are excluded. Horizontal spread within a row reflects trial-to-trial variance; vertical dispersion of row centres reflects task-to-task variance.}
\label{fig:ridges}
\end{figure}

The aggregate numbers in Sec.~\ref{sec:aggregate} come from a dataset with substantial trial-to-trial and task-to-task variance, both of which matter for how the figures should be read.
Figure~\ref{fig:ridges} plots the per-trial distributions of input and output tokens across the 27 tasks (excluding calibration): each row is a task and shows its ten cleaner-side and ten messier-side trials.
Spread within a row is trial-to-trial variance; dispersion of row centres across rows is task-to-task variance.
Both are visibly large.

\textbf{Trial-to-trial variance.}
Running the same prompt against the same codebase produces wildly different footprints.
Take the ten repetitions of a single task on a single side as a group.
On input tokens, the most expensive trial in a group typically costs around $2.5\times$ the cheapest, and roughly $72\%$ of groups have a ratio above $2\times$.
The most extreme example of this is \texttt{ckan/organization-list-exclude-empty}, where ten cleaner-side trials span $1.4$M to $10.6$M input tokens and ten messier-side trials span $1.8$M to $14.3$M.
Reasoning characters and files read show an even wider variance. 
The outlier filter described in Sec.~\ref{sec:setup} dampens this but does not erase it.
A single per-task delta of around $10\%$ may therefore be agent noise rather than signal; the dataset-level numbers in Sec.~\ref{sec:aggregate} hold because they pool hundreds of trials.

\textbf{Task-to-task spread.}
Different tasks disagree with each other on which side wins.
On \texttt{ckan/organization-list-exclude-empty}, the cleaner side uses $24\%$ fewer input tokens than the messier side; on \texttt{ckan/package-update-change-summary}, $44\%$ more.
Same repository, two tasks, opposite outcomes.
Across the 27 real-track tasks, per-task input-token deltas span $-47\%$ to $+44\%$ with a median of $-4.5\%$.
On 16 of 27 tasks, the agents working on the cleaner side use fewer input tokens, and on 11, they use more.
This is closer to 60/40 than the dataset-level $-7.1\%$ alone suggests.
Files read flips the direction entirely: on 10 tasks the agents on the cleaner side open fewer files, and on 17 they open more.
The aggregate findings hold up because they average over a wide spread of individual deltas, not because most tasks behave the same way.

\subsection{Case studies}
\label{sec:cases}

We outline a ``tension'' towards the end of~Sec.~\ref{sec:tracks}. To explain it better, we outline two tasks that exemplify this tension. Both come from the public pairs in our benchmark, and both involve a similar kind of cleanup: the cleanup pipeline acting on a god structure. They sit on opposite ends of the agent-footprint contrast.

\textbf{Case 1: (cleaner wins) \texttt{commons-bcel/stack-effect-annotation-in-disassembly}.}
The task asks the agent to add a stack-effect annotation to BCEL's bytecode disassembly output. On the messier variant, opcode dispatch lives inside two parallel god methods, each a several-hundred-line switch over JVM opcodes. The cleanup pipeline replaced both with thin dispatchers (a couple of dozen lines apiece) delegating to roughly ten named helpers each.

On this task agents working on the cleaner repo use $35\%$ fewer input tokens, open $25\%$ fewer files, edit $31\%$ fewer lines of code, and take $32\%$ fewer conversation turns to reach a solution. The files themselves are not smaller on the cleaner variant: \texttt{Utility.java} is essentially identical in size on both sides, and \texttt{CodeHTML.java} is in fact larger on the cleaner side because the extracted helpers carry their own boilerplate. 
Rather, because the agent can target (\texttt{grep}) precisely where the edits need to happen, it ends up navigating in a more token-efficient manner.

\textbf{Case 2 (messier wins): \texttt{genie/cluster-active-job-limit}.}
The task asks the agent to add a maximum-active-jobs cap to a Genie cluster's job-launch path. 
The cleanup pipeline on this part of the codebase extracted helpers around the focal launch logic but kept the original logic in place.
The file, therefore, is of roughly the same size on both variants. The cleaner side, however, has the surrounding work spread across more methods.
The effects of this show up as an $8\%$ increase in input tokens when working on the cleaner variant. 
The per-task deltas on every other metric we track sit within a few percent of zero (output tokens $-5\%$, reasoning $-4\%$, turns $+2\%$, files read $+1\%$, lines edited $-6\%$). %

The contrast with Case~1 is the Sec.~\ref{sec:tracks} tension at task level. On BCEL, the cleanup pipeline replaced the god dispatchers, and the agents could easily use grep to find relevant code, instead of scanning a 250-line switch. On this Genie task, the pipeline left the focal logic intact and added structure around it, and the agent paid for the added surface area without a corresponding navigability gain. %
We noticed similar patterns of behaviour on the private repositories as well. 

\subsection{Comment-volume ablation}
\label{sec:normalization}

The differences in agents' footprints, as we discussed above, came from pairs (of repositories) that differ on more than code. 
The processes of generating minimal pairs (Sec.~\ref{sec:minimal_pairs}) also move comments and suppression markers when they move code. 
As outlined in Table~\ref{tab:normalization}, consider that the cleaner version of \texttt{sonar-caas-poc} has $8000$ more comment lines than the messier version. 
The slopify process removed many long docstrings, verbose section headers, and explanatory blocks on private helpers.

Further, consider that comments are not inert to an agent. 
A one-line \texttt{\# keep in sync with metrics/launch\_rate.py} sitting above a function can redirect which files the agent touches. %
A repository whose comments carry such hints may therefore be accelerating the agent's exploration, deflating its footprint independently of code structure. 
On the other hand, low-information comments, such as banners, paraphrases of the line below, or stale TODOs, inflate the input-token count without meaningfully changing the agent's trajectory. 
So if we don't control for the comment gap, the findings above: "cleaner code costs less" might really be "code with fewer comment lines costs less". 

\begin{table}[h]
\centering
\caption{Per-pair effects of normalization on comment volume, suppression markers, and input-token footprint. Comment-line gap is reported as \emph{cleaner} $-$ \emph{messier} (positive $=$ cleaner side has more comment lines). ``Markers'' counts \texttt{\# noqa}, \texttt{// NOSONAR}, \texttt{\# TODO}, \texttt{\# FIXME}, and language-specific equivalents across both sides combined. $\Delta = (\text{cleaner} - \text{messier})/\text{messier}$ on input tokens; negative $=$ cleaner side is cheaper. Outlier filter as in Sec.~\ref{sec:setup}; $n{=}5$ trials per cell on the normalized pairs, against $n{=}10$ in Sec.~\ref{sec:results}.}
\label{tab:normalization}
\begin{tabular}{lccrr}
\toprule
Pair & \shortstack{Comment-line gap\\(orig.\ $\to$ norm.)} & \shortstack{Markers (total)\\(orig.\ $\to$ norm.)} & 
\shortstack{Input tok\\Orig.\ $\Delta$} & 
\shortstack{Input tok\\Norm.\ $\Delta$} \\ 
\midrule
\texttt{sonar-caas-poc}          & $+8{,}205 \to +223$ & $570 \to {\approx}50$ & $+1.2\%$  & $-18.0\%$ \\
\texttt{sonar-sca}               & $+1{,}281 \to -558$ & $221 \to 0$           & $+1.3\%$  & $-8.5\%$  \\
\texttt{sonarcloud-codedatalake} & $-258 \to -292$     & $542 \to {\approx}7$  & $-29.0\%$ & $-30.6\%$ \\
\texttt{genie}                   & $+54 \to +56$       & $197 \to 1$           & $+5.2\%$  & $+3.0\%$  \\
\bottomrule
\end{tabular}
\end{table}

This experiment aims to strip away that confounding factor by \textit{normalizing} the comments across the two variants of a minimal pair. We equalize comment content across the two sides of each pair. We remove the redundant comments on the heavier side (section banners, paraphrases of the line below, verbose docstrings on private helpers, dead debug remnants). Where a declaration with a byte-equivalent signature exists on both sides but only one side carries its docstring or Javadoc, we also restore the docstrings. 
Where comments differ due to underlying code, such as a Javadoc on a helper that exists only on the cleaner side, they are left in place.
Executable code is left untouched. 
We also strip suppression markers 
(\texttt{\# noqa}, 
 \texttt{// NOSONAR},
 \texttt{\# TODO},
 \texttt{\# FIXME}, 
and language-specific equivalents) from both sides. 

The aforementioned Table~\ref{tab:normalization} also outlines the effects of this normalization step. Two of our repositories, i.e., \texttt{sonarcloud-codedatalake} and \texttt{genie} already had a similar spread of comments between the cleaner and messier variants, but had a different set of markers. The before-and-after difference in agent footprints on these can be treated as solely caused by the markers. 

On both these repos, we find that agent footprint deltas do not change before and after normalization. 
In other words, agents spend similar amounts of input tokens on cleaner and messier versions of the code before and after normalization.
This tells us that suppression markers have negligible effects on agent trajectories. 
On \texttt{sonar-sca} and \texttt{sonar-caas-poc}, the comment asymmetry before normalization was substantial. On both, the cleaner side's advantage becomes more emphasized once that asymmetry is removed. 
\texttt{sonar-caas-poc} moves from $+1.2\%$ to $-18.0\%$, the largest shift in the table. 
Normalization here, however, drastically reduced the comments on the cleaner side, so some fraction of the $-18.0\%$ reflects the agent simply reading smaller files. 
To estimate a lower bound, we re-ran the comparison while controlling for LOC differences in the files the agent actually read. 
Even under that adjustment, (adjusted original $+2.9\%$, adjusted normalized $-13.0\%$, vs the raw $+1.2\% \to -18.0\%$ in Table~\ref{tab:normalization}), we find that post-normalization, the effect persists.

Based on this result, we conclude that comments and suppression markers are not what drives the cleaner-versus-messier footprint contrast. 
However, we do not read this as evidence that Sec.~\ref{sec:results} understated the effects of cleaner code. 
Our normalization process was an involved undertaking, with many manual interventions. 
The variability of this process, combined with the high variance of our results, as outlined in Sec.~\ref{sec:variance}, prevents us from making this stronger claim.

\section{Related Work}
\label{sec:related}

A substantial body of work exists on benchmarking the coding abilities of language models, and more recently, of coding agents built on top of them. We do not attempt an exhaustive review, and instead discuss the work this paper most directly builds on.

SWE-bench~\citep{jimenez2024swebench} and its filtered subset SWE-bench Verified \citep{chowdhury2024swebenchverified} established the task format that most current coding-agent evaluation uses: an underspecified natural-language instruction drawn from a real issue, a hidden test suite at the public surface, and a reference implementation kept internal to the benchmark. Each instance is one repository at one revision, so any difference an agent encounters across instances confounds task difficulty with code state. We borrow the task shape from these works, and add the within-pair controlled comparison that the format does not provide.

A more recent line of work argues that pass rate alone does not adequately characterise an agent's behaviour on a coding task, and additionally reports the resources the agent consumes in completing it. SWE-Effi~\citep{fan2025sweeffi} introduced this argument with resource-aware metrics that combine accuracy with token and wall-clock cost, and identified an ``expensive failures'' pattern in which unresolved attempts consume on average four times more resources than successful ones. Tokenomics~\citep{salim2026} applies a similar analysis across software-development-lifecycle phases. The authors dissect ChatDev~\citep{qian-etal-2024-chatdev} on GPT-5 and find that the code-review phase alone accounts for $59.4\%$ of token consumption, with input tokens making up $53.9\%$ of the total. AgentTaxo~\citep{wang2025agenttaxo} taxonomises multi-agent systems into planner, reasoner, and verifier roles. It defines the ``communication tax'' as duplicate tokens passed between roles, and reports a 2:1 to 3:1 input-to-output ratio across the systems it studies. 

The work closest to ours is~\citet{bai2026}, who measure token consumption on SWE-bench Verified across eight frontier LLMs and report four findings: agentic tasks consume roughly $1000\times$ the tokens of single-turn reasoning or chat workloads, input tokens dominate the total, per-task token usage is highly variable across repeated runs (up to $30\times$ on the same task), and accuracy peaks at intermediate token budgets rather than the largest. We corroborate two of those findings (input dominance, per-task variance) in our experiments. These papers vary properties of the agent (the model, the harness, the multi-agent topology, the lifecycle phase) while holding the codebase fixed. Our experiment does the converse: we vary the codebase while holding the agent and the task fixed, and the input-token dominance these papers report continues to hold under variation of the code.

Closer in spirit, two recent papers also vary the code while holding the model and the task fixed.
\citet{le2025names} construct minimal pairs of classes and methods from ClassEval~\citep{du2024classeval} and LiveCodeBench~\citep{jain2025livecodebench}, where the two sides differ only on identifier names.
They report sharp drops in summarization accuracy on the ClassEval pairs under uninformative renames (e.g., GPT-4o from $87.3\%$ to $58.7\%$), and smaller but persistent degradation on execution prediction despite behavior being preserved.
This echoes our Case~1 in Sec.~\ref{sec:cases}, where restoring named helpers gave the agent grep-targetable structure and produced the experiment's largest per-task footprint reduction.
\citet{simoes2025readability} construct minimal pairs of twelve Java classes drawn from OpenJDK~\citep{openjdk} along three independent axes: presence of comments, identifier semantics, and resolution of SonarQube-flagged issues.
They score readability across nine LLMs. They find that renaming identifiers is the most impactful, stripping comments less so, and resolving flagged issues barely affects the LLM.
Both studies build their minimal pairs at the file or snippet level, and probe them through single-turn LLM calls.
In contrast, our work experiments with widely-used multi-turn coding agents working over full repositories with hidden tests at the public surface.

SlopCodeBench~\citep{orlanski2026slopcodebench} is a long-horizon counterpart to our experiment. It comprises 36 problems, with each agent repeatedly extending its own prior code under evolving specifications, and tracks trajectory-level quality through verbosity and structural erosion.
The agent-generated code drifts generally toward more verbose, less structured output, ending $2.3\times$ longer than the human-maintained baselines.
Where we measure agent behaviour for a single task on a fixed codebase, SlopCodeBench measures the codebase an agent produces over many iterations. Our short-horizon results describe the upstream input to the iterative process they study, and the compounding question we discuss in Sec.~\ref{sec:limitations} is the setting they report on.

\section{Limitations}
\label{sec:limitations}

\textbf{Author-curated end to end.} Benchmarks like SWE-bench draw tasks from real GitHub issues, which keeps task selection at arm's length from the team measuring agent behaviour. Our setup has no such separation: we selected the repositories, built both sides of each pair, and authored every task that runs on them. The controlled comparison we wanted is only feasible under these conditions, but any systematic bias in our task selection would propagate directly to the findings we report.

\textbf{One configuration.} All numbers come from Claude Sonnet 4.6 inside Claude Code, on Python and Java codebases. A Haiku 4.5 sweep was too noisy on these tasks to extract footprint cleanly, and we ran neither GPT-family nor Gemini-family agents, nor harnesses other than Claude Code. The mechanism we identify (an agent's reread behaviour reacting to local code structure) plausibly transfers across models and harnesses, but the transfer is conjecture in this paper.

\textbf{Tokens, not dollars.} We use tokens as our cost proxy throughout. The dollar cost of an agent trajectory depends on the model, the provider, the cache state at the call, and queue delays, none of which are properties of the codebase. The token-to-dollar relationship is also nonlinear, and the reductions we report should map to dollar reductions of a different, configuration-dependent magnitude.

\textbf{Hidden tests only.} Pass rate scores the agent's final state against the hidden tests we wrote for each task. We do not check whether the agent broke unrelated tests already present in the repository, and a cleaner-side and messier-side solution that both pass the hidden test may still differ on tests they were not graded on.

\textbf{No post-facto cleanliness scan of agent output.} A natural follow-up question is whether agents working on cleaner code produce cleaner code in turn. We did not run SonarQube against the agent's final state, and cannot speak to whether cleanliness is self-reinforcing or self-eroding under agent work.

\textbf{``Cleaner'' and ``messier'' are within-pair labels.} They describe the two sides of a constructed pair, not a verdict on the original codebase. We do not claim that \texttt{commons-bcel}, \texttt{genie}, or \texttt{ckan} as shipped are messy repositories; only that our cleanup pipeline produced a variant of each that carries fewer SonarQube issues, and that this variant was the cleaner side of the pair in our experiments.

\textbf{Short-horizon.} The per-task footprint reductions we report are modest. The version of the maintainability argument worth measuring next is whether they compound: across a year of agent work on a codebase, do per-task savings accumulate, or does the codebase drift in ways that erase them? 
~\citet{orlanski2026slopcodebench} show us to evaluate an agent across an evolving set of tasks; porting our minimal-pair methodology will enable answering these questions.

\section{Conclusion}
\label{sec:conc}

A recurring claim about agent-driven software development is that maintainability principles were calibrated to human readers and lose their relevance as agents take over the work. Our experiment is a small piece of evidence against that claim. Across six minimal pairs and $33$ tasks, cleaner code did not change whether the agent finished its task, but it reduced the footprint the agent left doing it. The clearest behavioural marker was file revisitation: on the cleaner side, the agent re-opened already-edited files roughly a third less often. 
The principles that make code easier to maintain for a human developer appear to do the same for an agent.
However, our research leaves open the question of whether these per-task improvements compound over the long-term evolution of a codebase managed by agents.

\bibliography{references}
\bibliographystyle{iclr2026_conference}

\end{document}